\renewcommand{\r}{{\mathbb R}}
\newcommand{\hslambda}{{\cal H}^{\Lambda}}
\newcommand{\holambda}{H^{\Lambda}}
\newcommand{\Ulambda}{U^\Lambda}
\newcommand{\Uilambda}{U^{I,\Lambda}}
\newcommand{\bbbone}{\mathchoice {\rm 1\mskip-4mu l} {\rm 1\mskip-4mu l}
{\rm 1\mskip-4.5mu l} {\rm 1\mskip-5mu l}}
\newcommand{\Flambda}{{\cal F}^\Lambda}
\newcommand{\Hlambda}{{\cal H}^\Lambda}
\newcommand{\fer}[1]{(\ref{#1})}
\newcommand{\btau}{{\boldsymbol \tau}}
\newcommand{\bmu}{{\boldsymbol \mu}}
\newcommand{\blambda}{{\boldsymbol \lambda}}
\newcommand{\bQ}{{\boldsymbol Q}}
\newcommand{\av}[2]{{\left\langle{#1}\right\rangle_{\!\!{#2}}}}
\newcommand{\tr}{{\mbox{tr\,}}}
\newcommand{\bq}{{\boldsymbol q}}
\newcommand{\dbar}{{d\!\!\!^{\,\bf -}}\!\!}
\newcommand{\open}[1]{{{#1}^{{\!\!\!}^{{ }^\circ}}}}
\newcommand{\e}{{\cal E}\!}
\newcommand{\car}{{\mbox{CAR}}}

\documentstyle[12pt,amsmath,amssymb,epic]{article}

\begin{document}

\title{\bf Statistical Mechanics of Thermodynamic Processes\footnote{Submitted
 for publication to World Scientific Publishing Co.}}

\author{
J. Fr\"{o}hlich$^{\rm 1}$\footnote{juerg@itp.phys.ethz.ch}\ ,\  M. 
Merkli$^{\rm 1}$\footnote{merkli@itp.phys.ethz.ch}\ ,\   S. Schwarz$^{\rm
  1}$\footnote{sschwarz@itp.phys.ethz.ch}\ ,\  D. Ueltschi$^{\rm
2}$\footnote{ueltschi@math.ucdavis.edu}\\
\ \\
$^{\rm 1}$Theoretische Physik \\
 ETH-H\"{o}nggerberg \\ 
CH-8093 Z\"{u}rich, Switzerland
\vspace*{.2cm}
\\
$^{\rm 2}$
Department of Mathematics\\
 University of California\\
Davis, CA 95616, USA }
\date{\today}
\maketitle

\centerline{
This note is dedicated to H. Ezawa on the occasion of his $70^{\rm th}$
birthday,}
\centerline{ with respect and affection.}

\section{Time-dependent thermodynamic processes}     %S1-Heads

In this note we describe some results concerning non-relativistic quantum
systems at positive temperature and density confined to macroscopically large
regions, $\Lambda$, of physical space $\r^3$ which are under 
the influence of some local, time-dependent external forces.  We are
interested in asymptotic 
properties of such systems, as $\Lambda$ increases to all of ${\mathbb
  R}^3$. It might thus appear natural to directly study such systems in the
thermodynamic limit, $\Lambda\nearrow \r^3$. But for reasons of
technical simplicity and ease of exposition we prefer to first consider finite
systems and then extend our results to the thermodynamic limit. An important
reference is \cite{2}. Details of our results appear in \cite{1,4,5}.

The Hilbert space of pure state vectors of a system confined to $\Lambda$ is
denoted by $\hslambda$, and its dynamics is generated by a {\it time-dependent}
Hamiltonian $\holambda_t$ with the properties that $\holambda_t$ is a
selfadjoint operator on $\hslambda$, for each time $t$, that its domain of
definition is time-independent, and that
$\dot{H}_t^\Lambda=\frac{d}{dt}\holambda_t$ is bounded by $\holambda_s$,
e.g. in the sense of Kato-Rellich \cite{8}, for arbitrary times $t$ and
$s$. If the sytem is in a state corresponding to a vector $\psi_s\in\hslambda$,
at time $s$, then its state vector, $\psi_t$, at time $t$ is given by
\begin{equation}
\psi_t=\Ulambda(t,s)\psi_s,
\label{1}
\end{equation}
where $\Ulambda(t,s)$ denotes the {\it unitary propagator} on
  $\hslambda$. This operator is the solution of the equation
\begin{equation}
\frac{\partial}{\partial t} \Ulambda(t,s)=-i \holambda_t\Ulambda(t,s)
\label{2}
\end{equation}
with
\begin{equation}
\Ulambda(s,s)=\bbbone
\label{3}
\end{equation}
and has the property that
\begin{equation}
\Ulambda(t,s)=\Ulambda(t,u) \Ulambda(u,s),
\label{4}
\end{equation}
for arbitrary pairs $(t,s), (t,u)$ and $(u,s)$ of times. (We are using units
such that Planck's constant $\hbar=1$.) \\
\indent
The kinematics of the system is encoded in an algebra $\Flambda$ of bounded
operators on $\Hlambda$ with the properties that 
\begin{equation}
\Flambda\subseteq {\cal B}(\Hlambda),
\label{m3}
\end{equation}
and 
\begin{equation}
\Ulambda(s,t) A \Ulambda(t,s)\in \Flambda,
\label{m1}
\end{equation}
for all $A\in\Flambda$ and all times $s,t$. \\
\indent
The time evolution of a time-dependent family of operators
\begin{equation*}
\{A_t\}_{t\in\r}\subset \Flambda
\end{equation*}
in the {\it Heisenberg picture} is given by
\begin{equation}
A(t):= \Ulambda(t_0,t) A_t \Ulambda(t,t_0)
\label{5}
\end{equation}
where $t_0$ denotes the ``initial time''; (e.g., the time when an experiment
involving the system is started). We also denote the r.h.s. of \fer{5} by 
\begin{equation}
\alpha_{t_0,t}^\Lambda(A_t).
\label{6}
\end{equation}
Then
\begin{equation}
A(t)=\alpha_{t_0,t}^\Lambda(A_t)=\alpha_{t_0,t_1}^\Lambda\circ\alpha_{t_1,t}^\Lambda(A_t),
\label{7}
\end{equation}
for an arbitrary time $t_1$.  One easily verifies that 
\begin{equation}
\frac{d}{dt} A(t)=\alpha^\Lambda_{t_0,t}\left(\frac{DA_t}{Dt}\right),
\label{8}
\end{equation}
where the {\it Heisenberg derivative}, $DA_t/Dt$, is defined by
\begin{equation}
\frac{DA_t}{Dt}=i [H_t^\Lambda,A_t]+\dot A_t.
\label{9}
\end{equation}

We assume that the Hamiltonians $\holambda_t$ are of the form
\begin{equation}
\holambda_t= \holambda_0+W_t,
\label{23}
\end{equation}
where the term $W_t$ describes a time-dependent perturbation of the
system. When this perturbation is turned off the propagator is given by the
unitary group 
$\{e^{it\holambda_0}\}_{t\in\r}$ on $\hslambda$ implementing the Heisenberg
time evolution 
\begin{equation}
\alpha_t^{0,\Lambda}(A_t):=e^{it\holambda_0} A_t e^{-it\holambda_0}.
\label{m2}
\end{equation}

The unperturbed system may exhibit a group of dynamical internal symmetries
unitarily represented on $\hslambda$. For
reasons of simplicity of our exposition, we assume that the symmetry group is
a connected compact Lie group $\cal G$. Let $\cal Z$ denote an $n$-dimensional
continuous connected subgroup contained in or equal to the centre of the group
$\cal G$, and let $Q_1^\Lambda,\ldots,Q_n^\Lambda$ denote the generators of the
unitray representation of $\cal Z$ on $\hslambda$. The operators
$Q^\Lambda_1,\ldots Q_n^\Lambda$ are selfadjoint operators on $\hslambda$ with
\begin{equation}
[Q_i^\Lambda,Q_j^\Lambda]=0, \mbox{\ \ for all $i,j=1,\ldots,n$},  
\label{10}
\end{equation}
(in the sense that their spectral projections commute), and, since $\cal G$ has
been assumed to be a group of {\it dynamical} symmetries, 
\begin{equation}
e^{i\btau\cdot\bQ^\Lambda}
e^{it\holambda_0}=e^{it\holambda_0}e^{i\btau\cdot\bQ^\Lambda},
\label{11}
\end{equation}
for arbitrary $\btau=(\tau_1,\ldots,\tau_n)$ and arbitrary $t$; (here
$\btau\cdot\bQ^\Lambda:= \sum_{j=1}^n \tau_jQ_j^\Lambda$). We define {\it
  gauge transformations of the first kind\,} by
\begin{equation}
\phi_{\btau}^\Lambda(A):= e^{i\btau\cdot\bQ^\Lambda} A
e^{-i\tau\cdot\bQ^\Lambda},
\label{12}
\end{equation}
for arbitrary $A\in\Flambda$. It is assumed that $\phi_\btau^\Lambda$
are $*$automorphisms of $\Flambda$. We define the $C^*$-algebra ${\cal
  A}^\Lambda$ of ``observables'' to be the
fixed-point subalgebra of the algebra $\Flambda$ with respect to the
automorphism 
group $\{\phi^\Lambda_\btau\}_{\btau\in\r^n}$; i.e., 
\begin{equation}
{\cal A}^\Lambda:=\{ A\in\Flambda\ |\ \phi_\btau^\Lambda(A)=A,\
\forall\btau\}.
\label{13}
\end{equation}

{\it Mixed states} of the system are described by {\it density matrices},
$\varrho$, on $\hslambda$, (i.e., by non-negative trace-class operators with
tr$\varrho=1$). If the perturbation $W_t$ of the system vanishes, i.e.,  
\begin{equation}
\holambda_t=\holambda_0, \mbox{\ \ \ for all times $t$,}
\label{14}
\end{equation}
then the notion of {\it thermal equilibrium} of the system is meaningful. At
  {\it 
  inverse temperature $\beta$} and {\it chemical potentials
  $\mu_1,\ldots,\mu_n$}, the equilibrium state is given by the density matrix
\begin{equation}
\varrho_{\beta,\bmu}:=\left(\Xi_{\beta,\bmu}^\Lambda\right)^{-1} \exp
-\beta[\holambda_0-\bmu\cdot\bQ^\Lambda]. 
\label{15}
\end{equation}
It is assumed, here, that $\exp -\beta[\holambda_0-\bmu\cdot\bQ^\Lambda]$
is trace-class, for arbitrary $\beta>0$, $\bmu\in\r^n$; the normalization
factor $\Xi_{\beta,\bmu}^\Lambda$, the so-called {\it grand partition
  function}, is chosen such that tr$\varrho_{\beta,\bmu}=1$, and one commonly
assumes that the system is {\it thermodynamically stable}, in the sense that
the thermodynamic potential, $G^\Lambda$, given by 
\begin{equation}
\beta G^\Lambda(\beta,\bmu):=-\ln \Xi_{\beta,\bmu}^\Lambda
\label{16}
\end{equation}
is {\it extensive}, i.e., bounded in absolute value by a constant times the
volume of $\Lambda$, for arbitrary $\beta>0, \bmu\in\r^n$, and
$\Lambda\nearrow\r^3$. \\
\indent
If, at time $t_0$, the system is in a mixed state $\varrho(t_0)$ then its
state at time $t$ is given by the density matrix
\begin{equation}
\varrho(t)=\Ulambda(t,t_0) \varrho(t_0)
\Ulambda(t_0,t)=\alpha_{t,t_0}^\Lambda(\varrho(t_0)).
\label{17}
\end{equation}
Then, using equations \fer{5} and \fer{17}, we find that  
\begin{equation}
\av{A_t}{\varrho(t)}:=\mbox{tr}\left(\varrho(t)A_t\right)=\mbox{tr}\left(\varrho(t_0)A(t)\right)=:\av{A(t)}{\varrho(t_0)},
\label{18}
\end{equation}
as expected. \\
\indent
The {\it entropy} of a state given by a density matrix $\varrho$ is
defined by 
\begin{equation}
S(\varrho)=-\mbox{tr}\left(\varrho\ln\varrho\right).
\label{19}
\end{equation}
(We use units such that Boltzmann's constant $k_B=1$.)\\
\indent
Since $\Ulambda(t,t_0)$ is unitary, for arbitrary $t, t_0$, it follows from
\fer{17} and the cyclicity of the trace that
\begin{equation}
S(\varrho(t))=S(\varrho(t_0)),
\label{20}
\end{equation}
for arbitrary $t,t_0$.\\
\indent
Next, we introduce the notion of a (time-dependent) {\it thermodynamic
  process}. We imagine that, for all times $t\leq t_0$, the Hamiltonian
$\holambda_t=\holambda_{t_0}=:\holambda_0$ is {\it independent} of time $t$,
and that the initial state at time $t_0$ of the system is given by an {\it
  equilibrium state} $\varrho_{\beta,\bmu}=:\varrho_{\beta,\bmu}(t_0)$, as
defined in equation \fer{15}, for some inverse temperature $\beta$ and
chemical potentials $\bmu$. We are interested in studying the effects of {\it
  local, external perturbations} acting on the system. In order to make more
precise what we are talking about, we assume that the systems considered in
this note have a {\it local structure}: if $\Lambda_0$ is an arbitrary convex
subset of the convex region $\Lambda$ containing the system, and
$\Lambda\backslash\Lambda_0$ denotes its complement then the Hilbert space
$\hslambda$ of the system can be factorized into
\begin{equation*}
\hslambda={\cal H}^{\Lambda_0}\otimes{\cal H}^{\Lambda\backslash\Lambda_0},
\end{equation*}
where ${\cal H}^{\Lambda_0}$ can be interpreted as the Hilbert space of pure
state vectors of the degrees of freedom localized in $\Lambda_0$. Let ${\cal
  F}^{\Lambda_0}\subseteq {\cal B}({\cal H}^{\Lambda_0})$ be the kinematical
algebra associated to the region $\Lambda_0$, see \fer{m3}. Then the subalgebra
\begin{equation}
{\cal F}^{\Lambda_0}\otimes\bbbone|_{{\cal
    H}^{\Lambda\backslash\Lambda_0}}\subset\Flambda
\label{21}
\end{equation}
is naturally identified with ${\cal F}^{\Lambda_0}$. Without any essential
loss of generality, we may assume that the gauge transformations
$\phi_\btau^\Lambda$ introduced in equation \fer{12} leave the subalgebra
${\cal F}^{\Lambda_0}$ of $\Flambda$ invariant, and when restricted to ${\cal
  F}^{\Lambda_0}$ coincide with $\phi_\btau^{\Lambda_0}\otimes\mbox{id}|_{{\rm
    Aut}({\cal F}^{\Lambda\backslash\Lambda_0})}$. Then the algebra ${\cal A}^{\Lambda_0}$
can be identified with
\begin{equation}
\{ A\in{\cal F}^{\Lambda_0}\ | \ \phi_\btau^\Lambda(A)=A\}
\label{22}
\end{equation}
and will be viewed as a subalgebra of ${\cal A}^\Lambda$, for arbitrary
$\Lambda_0\subset\Lambda$. \\
\indent
In the following, we shall keep $\Lambda_0\subset\Lambda$ fixed and view the
degrees of freedom localized in $\Lambda_0$ as a finite subsystem of the
entire system, while the regions $\Lambda$ will be let to increase to $\r^3$,
eventually. 

The interaction term $W_t$, given in \fer{23}, describes the dynamical effects
of an external perturbation acting on the system and is assumed to have the
following properties:
\begin{itemize}
\item[(i)] $W_t=0$, for $t<t_0$; and
\item[(ii)] $W_t$ is {\it local} in the sense that $W_t\in {\cal
    F}^{\Lambda_0}$, for all times $t$, where $\Lambda_0$ is an arbitrary, but
  fixed bounded, convex subset of $\r^3$ ({\it independent} of $t$).  
\end{itemize}
A thermodynamic process is {\it charge-conserving} iff $W_t$ is {\it
  gauge-invariant}, i.e., $W_t\in{\cal A}^{\Lambda_0}$, for all times $t$. \\
\indent
Later, we shall also assume that $W_t$ is {\it small} in the sense that a
  suitable norm of $W_t$ is assumed to be small, {\it uniformly} in $t$.\\
\indent
The perturbation $W_t$ may describe, for example, the effects of shining a
  focussed beam of light into the system, or of local, time-dependent
  variations of an external magnetic field applied to the system, or of the
  motion of a piston confining particles to a time-dependent
  subset, $\Lambda_t$, of $\Lambda$, with $\Lambda\backslash\Lambda_t\subseteq
  \Lambda_0$. Thus, $W_t$ is typically of the form
\begin{equation}
W_t=W(\blambda(t)),
\label{24}
\end{equation}
where $\blambda=(\lambda_1,\ldots,\lambda_k)$ is a finite set of {\it external
  control parameters}, and the time-dependence of $W_t$ is entirely due to a
  possible time-dependence of the control parameters $\blambda$. \\
\indent
If the ratio volume$(\Lambda_0)$:volume$(\Lambda)$ is very small the subsystem
  in the region $\Lambda\backslash\Lambda_0$ can be interpreted as a {\it
  thermostat} for the small subsystem in $\Lambda_0$, keeping the values of
  the temperature and the chemical potentials constant throughout a
  thermodynamic process. Since we have assumed that the initial state,
  $\varrho(t_0)=\varrho_{t_0}$, of the entire system at time $t_0$ is an
  equilibrium state, 
\begin{equation}
\varrho(t_0)=\varrho_{t_0}=\varrho_{\beta,\bmu},
\label{25}
\end{equation}
as defined in equation \fer{15}, with $\holambda_{t_0}=\holambda_0$, we are
studying thermodynamic processes at constant temperature $\beta^{-1}$ and
constant chemical potentials $\bmu$; (at least after passing to the
thermodynamic 
limit $\Lambda\nearrow\r^3$, with $\Lambda_0$ kept fixed).\\
\indent
The {\it true state} of the system at time $t$ is given by
\begin{equation}
\varrho(t)=\alpha_{t,t_0}^\Lambda(\varrho(t_0)),
\label{26}
\end{equation}
see equation \fer{17}. If the time-dependence of the perturbation
$W_t=W(\blambda(t))$ is {\it slow} it is of interest to compare the true
state $\varrho(t)$ of the system with a reference state,
$\varrho_t$, given by 
\begin{equation}
\varrho_t:=e^{\beta G^\Lambda(\beta,\bmu;\blambda(t))} \exp -\beta\left[
  \holambda_t-\bmu\cdot\bQ^\Lambda\right],
\label{27}
\end{equation}
where
\begin{equation}
\beta G^\Lambda(\beta,\bmu;\blambda(t))= -\ln \mbox{tr}\left( \exp -\beta\left[
    \holambda_t -\bmu\cdot\bQ^\Lambda\right]\right),
\label{28}
\end{equation}
and $\blambda(t)$ are the time-dependent control parameters that give rise to
the time-dependence of $\holambda_t=\holambda_0+W(\blambda(t))$. We shall call
the state $\varrho_t$ in equation \fer{27} the {\it reference state} at time
$t$. \\
\indent
An important quantity in the characterization of thermodynamic processes is
the {\it relative entropy} of the reference state $\varrho_t$ with respect to
the true state $\varrho(t)$ of the system, which is given by 
\begin{equation}
S(\varrho_t|\varrho(t)):=-\tr\big(\varrho(t)[\ln\varrho_t-\ln\varrho(t)]\big)
=-\tr\left(\varrho(t)\ln\varrho_t\right)-S(\varrho(t)),
\label{29}
\end{equation}
where the entropy $S(\varrho(t))$ of $\varrho(t)$ has been defined in
\fer{19}. If $A$ is a non-negative trace-class operator and $B$ is a strictly
positive trace-class operator then 
\begin{equation*}
-\tr\left(A\ln B-A\ln A\right)\geq \tr\left(A-B\right);
\end{equation*}
see e.g. Lemma 6.2.21 of \cite{2}. Setting $A=\varrho(t)$ and $B=\varrho_t$,
and using that $\tr\varrho(t)=\tr\varrho_t=1$, we conclude that 
\begin{equation}
S(\varrho_t|\varrho(t))\geq 0,
\label{30}
\end{equation}
for all times $t$. By equation \fer{20},
\begin{equation}
S(\varrho(t))=S(\varrho(t_0))=:S(t_0).
\label{31}
\end{equation}
It then follows from \fer{29} and \fer{30} that 
\begin{equation}
S(t):=-\tr\big(\varrho(t)\ln\varrho_t\big)\geq S(t_0),
\label{32}
\end{equation}
for all times $t\geq t_0$. Next, we note that, by equation \fer{27},
\begin{equation}
S(t)=\beta\left[ \av{\holambda_t-\bmu\cdot\bQ^\Lambda}{\varrho(t)}
  -G(\beta,\bmu;\blambda(t))\right].
\label{33}
\end{equation}
It is natural to define the {\it internal energy}, $\Ulambda(t)$, of the sytem
at time $t$ by
\begin{equation}
\Ulambda(t):=\av{\holambda_t}{\varrho(t)},
\label{34}
\end{equation}
and the various {\it charge densities} by 
\begin{equation}
q_j^\Lambda(t):=\av{Q^\Lambda_j}{\varrho(t)}.
\label{35}
\end{equation}
Dropping superscripts $\Lambda$, it follows that 
\begin{equation}
G=U-\bmu\cdot\bq -TS,
\label{36}
\end{equation}
which is the usual relation between the Gibbs potential $G$ and the internal
energy, charge densities and the entropy. All these quantities are {\it
  extensive} and, hence, do not have a limit, as $\Lambda$ increases to
$\r^3$. It is more useful to consider their {\it time derivatives}. Taking the
time derivative of equation \fer{33} it follows that 
\begin{equation}
\dot S(t)=\beta\dot U(t) -\beta\bmu\cdot\dot\bq(t)-\beta\frac{\partial
  G}{\partial\blambda}\cdot\dot\blambda(t).
\label{37}
\end{equation}
The combination
\begin{equation}
\dbar A:=-\bmu\cdot d\bq -\frac{\partial G}{\partial\blambda}\cdot d\blambda
\label{38}
\end{equation}
is commonly interpreted as the {\it work} done by the system during a change
of state. Hence we conclude that
\begin{equation}
\dot U=T\dot S-\frac{\dbar A}{dt},
\label{39}
\end{equation}
which summarizes the first and second law of thermodynamics (for {\it
  reversible processes}).\\
\indent
It is important to notice that, under our assumptions on the perturbation
  operator $W_t=W(\blambda(t))$, the quantities 
\begin{equation*}
\dot U^\Lambda(t), \dot
  \bq^\Lambda(t) \mbox{\ \ and\ \ } (\partial
  G^\Lambda/\partial\blambda)\cdot\dot \blambda(t)
\end{equation*}
 have finite thermodynamic limits. By equations \fer{26} and
  \fer{34}, and because
\begin{equation*}
\frac{D\holambda_t}{Dt}=\dot H^\Lambda_t=\dot W_t=\frac{\partial
  W(\blambda(t))}{\partial\blambda}\cdot\dot\blambda(t)
\end{equation*}
it follows that 
\begin{equation}
\dot U^\Lambda(t)=\av{\frac{\partial
    W(\blambda(t))}{\partial\blambda}}{\varrho(t)}\cdot\dot\blambda(t).
\label{40}
\end{equation}
Similarly, 
\begin{eqnarray}
\dot\bq^\Lambda(t)&=&\av{\frac{D\bQ^\Lambda}{Dt}}{\varrho(t)}\nonumber\\
&=&i\av{[W(\blambda(t)),\bQ^\Lambda]}{\varrho(t)}
=-\frac{\partial}{\partial\btau}\av{\phi^\Lambda_\btau\Big(W(\blambda(t))\Big)}{\varrho(t)}\Big|_{\btau=0},
\label{41}
\end{eqnarray}
where the gauge transformations $\phi_\btau^\Lambda$ have been defined in
equation \fer{12}. Under our hypotheses on $W(\blambda(t))$,
the operators $\frac{\partial W(\blambda(t))}{\partial\blambda}$ and
$\frac{\partial\phi_\btau^\Lambda(W(\blambda(t)))}{\partial\btau}$ are
strictly {\it local}, in the sense that they are elements of the algebra ${\cal
  F}^{\Lambda_0}\subset \Flambda$, see \fer{21}, and {\it independent} of
$\Lambda$. In the next section, we shall see that the propagators in the {\it
  interaction picture}
\begin{equation*}
\Uilambda(t,s):=e^{it\holambda_0}\Ulambda(t,s)e^{-is\holambda_0}
%\label{42}
\end{equation*}
have thermodynamic limits, as $\Lambda\nearrow\r^3$, under standard
assumptions on the unperturbed evolution generated by $\holambda_0$. If
expectations of local, bounded operators in the initial state,
$\varrho_{t_0}=\varrho_{\beta,\bmu}$, of the system have thermodynamic limits,
as $\Lambda\nearrow\r^3$, which is a standard assumption (or result -- see
\cite{2} for examples -- ) then it follows that the thermodynamic limit of the
quantities on the r.h.s. of equations \fer{40} and \fer{41} exist. \\
\indent
Finally, from the definition of $G^\Lambda$, equation \fer{28}, and using the
cyclicity of the trace, we find that 
\begin{equation*}
\frac{\partial
  G^\Lambda}{\partial\blambda}(\beta,\bmu;\blambda(t))\cdot\dot\blambda(t)= \av{\dot
  H_t^\Lambda}{\varrho_t}=\av{\frac{\partial
  W(\blambda(t))}{\partial\blambda}}{\varrho_t}\cdot\dot\blambda(t),
\end{equation*}
hence
\begin{equation}
\frac{\partial G}{\partial\blambda}(\beta,\mu;\blambda(t))=\av{\frac{\partial
    W(\blambda(t))}{\partial\blambda}}{\varrho_t},
\label{43}
\end{equation}
which, under the same standard assumptions, has a well defined thermodynamic
limit. We have thus proven the equation 
\begin{eqnarray}
\dot S(t)&=&\beta \left[\av{\frac{\partial
      W(\blambda(t))}{\partial\blambda}}{\varrho(t)}-\av{\frac{\partial
      W(\blambda(t))}{\partial
      \blambda}}{\varrho_t}\right]\cdot\dot\blambda(t)\nonumber\\
&&
      +\beta\bmu\cdot\frac{\partial}{\partial\btau}\av{\phi_\btau^\Lambda\Big(W(\blambda(t))\Big)}{\varrho(t)}\Big|_{\btau=0}
\label{44}
\end{eqnarray}
for the rate of change in time of the entropy $S(t)$, and we have convinced
ourselves 
that all three terms on the r.h.s. of equation \fer{44} have well defined
thermodynamic limits. \\
\indent
Returning to equations \fer{29}, \fer{30} and \fer{32}, one may ask under what
conditions the inequalities in \fer{30} and \fer{32} are saturated. The answer
is given in Lemma 6.2.21 of \cite{2}:
\begin{equation}
S(\varrho_t|\varrho(t))=0 \mbox{\ \ \ \ iff \ \ \ $\varrho_t=\varrho(t)$},
\label{45}
\end{equation}
i.e., iff the reference state $\varrho_t$ coincides with the true state
$\varrho(t)$. In view of equation \fer{44} one may expect that a similar
result holds in the thermodynamic limit. Equation \fer{44} proves that the
thermodynamic limit of 
\begin{equation}
\Delta S(t):=S(t)-S(t_0)=\int_{t_0}^t \dot S(t') dt'
\label{46}
\end{equation}
exists, and one expects that if $\Delta S(t)=0$ then the true state of the
system at time $t$ is given by the thermodynamic limit of the reference states
$\varrho_t$; (see \cite{5}). Furthermore, if, at time $t$, the restrictions
of the true and the reference state to the subalgebra ${\cal F}^{\Lambda_0}$
coincide in the thermodynamic limit then $\Delta\dot S(t)=\dot S(t)=0$, as
follows from equation \fer{44}. \\
\indent
Of course, if $W_t$ depends non-trivially on time $t$, for $t>t_0$, there is
no reason why the true and the reference state should ever coincide at times
$t>t_0$. Then a relevant and interesting problem is to study the rate of
change of the entropy under the assumption that the perturbation $W_t$ depends
slowly on time, 
\begin{equation*}
W_t=W(\blambda(t/T)),
\end{equation*}
for some large $T$. In this situation one would like to prove an {\it
  adiabatic theorem} yielding sharp estimates on the rate at which $\dot S(t)$
  tends to $0$, as $T\rightarrow\infty$. More interestingly, such a theorem
  would tell us at which rate the differences of expectation values of local
  operators in the true state and in the reference state tend to $0$, as
  $T\rightarrow\infty$. In this note we shall not address this problem.\\
\indent
The essential features of our definition of the entropy $S(t)$ can be
  summarized as follows:\\
\indent
(1)\ It is compatible with the first and second law of thermodynamics; see
  equations \fer{37}-\fer{39}.\\
\indent
(2)\ $S(t)\geq S(t_0)$, for all $t\geq t_0$, i.e., entropy tends to
  increase.\\
\indent
(3)\ The thermodynamic limit of $\dot S(t)$, and hence of $\Delta
  S(t)=S(t)-S(t_0)$ exists, for all times $t$. \\
\indent
(4)\ If the time evolution is {\it adiabatic}, in the sense that the
  norm of the difference of the restrictions of the true and of the reference
  state to the algebra ${\cal F}^{\Lambda_0}$ is bounded by some small,
  positive number $\epsilon>0$, for all times $t\leq t_1$, then $|\dot
  S(t)|<O(\epsilon)$ and $\Delta S(t)<O(\epsilon)$, for $t\leq t_1$, i.e., the
  entropy remains approximately constant. For more detail, see \cite{4}.\\
\indent
In the following, we consider two typical examples of time-dependent
thermodynamic processes.
\begin{enumerate}
\item[] {\bf Process (I)}.\ The perturbation $W_t$ converges to a limiting
  operator $W_\infty\neq 0$, as $t\rightarrow\infty$, with 
\begin{equation}
\int^\infty  \|W_t-W_\infty\| dt<\infty.
\label{47}
\end{equation}
For such perturbations we study the phenomenon of {\it return to equilibrium}:
Under suitable assumptions on the unperturbed dynamics in the thermodynamic
limit (``dispersiveness'') and assuming that a suitable norm of $W_\infty$ is
small enough, we show that, in the thermodynamic limit, the true state of the
system converges to an equilibrium state w.r.t. the dynamics determined by
$H_\infty^\Lambda=H_0^\Lambda+ W_\infty$ at temperature $\beta^{-1}$, as $t\rightarrow\infty$, if the initial
state is an equilibrium state of the unperturbed dynamics at temperature
$\beta^{-1}$ or a local perturbation thereof.\\
\indent
For earlier results, see \cite{R,jp,1}. Our analysis is an extension of
results in \cite{H,R,3} and is based on methods developed in \cite{5}.\\
The result described here is a kind of adiabatic theorem and shows that
thermodynamic processes with perturbations $W_t$ as specified above are {\it
  reversible}. It implies that, in the thermodynamic limit,
\begin{equation}
\lim_{t\rightarrow\infty} \dot S(t)=0,\mbox{\ \ \ and hence\ \ \
  }\lim_{t\rightarrow\infty}\lim_{\Lambda\nearrow\r^3} \dot\bq^\Lambda(t)=0,
\label{48}
\end{equation}
i.e., the entropy production rate and the rate of change of the charges
$\bq^\Lambda(t)$ vanish in the thermodynamic limit, as time tends to
infinity.\\
 
\item[] {\bf Process (II)}.\ For $t\geq t_0$, the perturbation $W_t$ depends {\it periodically}
  on time $t$, with period $T$, see \cite{A}, and \cite{B1,B2} for recent
  experiments involving time-periodic perturbations. Under the same assumptions on the unperturbed
  dynamics in the thermodynamic limit as in (I) and if a suitable norm of
  $W_t$ is small enough, for all $t\in[t_0,t_0+T]$, we prove that, in the
  thermodynamic limit, the true state of the system converges to a {\it
    time-periodic state} of period $T$, as time $t$ tends to infinity.\\
It is not hard to generalize this to perturbations $W_t$ with the property
that  
\begin{equation}
\int^\infty  \|W_t-W_t^\infty\| dt<\infty,
\label{49}
\end{equation}
where $W_t^\infty$ is periodic in $t$ with some period $T$.
\end{enumerate}

\section{Processes (I) and (II) in the thermodynamic limit}

In this section we study the thermodynamic limit, $\Lambda\nearrow\r^3$,
  of thermodynamic processes, in particular of processes (I) and (II)
  described at the end of Section 1. It is convenient to introduce a
  $C^*$-algebra, ${\cal F}$, of operators for the infinite system. We define
\begin{equation}
\open{{\cal F}}=\bigvee_{\Lambda\nearrow\r^3}\Flambda
\label{50}
\end{equation}
to be the algebra generated by all the algebras $\Flambda$, for an increasing
sequence of bounded convex regions $\Lambda\nearrow\r^3$. The $C^*$-algebra
$\cal F$ is defined as the closure of $\open{{\cal F}}$ in the operator
norm. For an operator $A\in\Flambda$ and a set $\Lambda'\supseteq\Lambda$, one
can define
\begin{equation}
\alpha_t^{0,\Lambda'}(A):= e^{it{H}_0^{\Lambda'}} A e^{-it{H}_0^{\Lambda'}}. 
\label{m4}
\end{equation}
It is a standard assumption (that can be verified in physically relevant
examples -- see Section 3) that the norm-limit
\begin{equation}
n-\lim_{\Lambda'\nearrow\r^3}\alpha_t^{0,\Lambda'}(A)=:\alpha_t^0(A)
\label{51}
\end{equation}
exists, for all $A\in\Flambda$, for an arbitrary bounded convex set
$\Lambda\subset\r^3$. Since $\alpha_t^{0,\Lambda'}(A)$ belongs to ${\cal
  F}^{\Lambda'}$, it follows that $\alpha_t^0(A)\in{\cal F}$. By continuity,
$\alpha_t^0$ can be extended to a $*$automorphism group of the $C^*$-algebra
${\cal F}$.\\
\indent
For a finite system confined to a region $\Lambda\supset\Lambda_0$, we define
the {\it propagator in the interaction picture} by
\begin{equation}
\Uilambda(t,s):=e^{it\holambda_0} \Ulambda(t,s) e^{-is\holambda_0}.
\label{52}
\end{equation}
Assuming that the perturbation $W_t$ is norm-continuous in $t$, we can expand
$\Uilambda(t,s)$ in a Dyson series that converges in norm, uniformly in
$\Lambda$, and with the property that all terms in the series have a
thermodynamic limit. Setting
\begin{equation*}
W^{I,\Lambda}_t:=\alpha_t^{0,\Lambda}(W_t),
\end{equation*}
and 
\begin{equation}
W_t^I:=\alpha_t^0(W_t),
\label{53}
\end{equation}
we find that 
\begin{equation*}
\Uilambda(t,s)=\bbbone+\sum_{n\geq 1}(-i)^n\int_s^t dt_1\cdots\int_s^{t_{n-1}}
dt_n\  W^{I,\Lambda}_{t_1}\cdots W_{t_n}^{I,\Lambda}
\end{equation*}
and this operator converges in norm to 
\begin{equation}
U^I(t,s)=\bbbone+\sum_{n\geq 1} (-i)^n\int_s^tdt_1\cdots\int_s^{t_{n-1}} dt_n
\ W_{t_1}^I\cdots W_{t_n}^I,
\label{54}
\end{equation}
as $\Lambda\nearrow\r^3$. The propagator $U^I(t,s)$ solves the differential
equation 
\begin{equation}
\begin{array}{c}
\frac{\partial}{\partial t} U^I(t,s) =-i W_t^I U^I(t,s),  \\
U^I(s,s)=\bbbone 
\end{array}
\label{55}
\end{equation}
and $U^I(t,s)\in{\cal F}$, for all finite times $t,s$.\\
\indent
These remarks enable us to define the time evolution of an operator $A\in{\cal
  F}$ in the Heisenberg picture from time $s$ to time $t$ by
\begin{equation}
\alpha_{s,t}(A)=n-\lim_{\Lambda\nearrow\r^3}\alpha_{s,t}^\Lambda (A)=\alpha_{-s}^0\left(
  U^I(s,t)\alpha_t^0(A) U^I(t,s)\right);
\label{56}
\end{equation}
see equations \fer{52} and \fer{6}. Equation \fer{56} shows that
$\alpha_{s,t}$ is a $*$automorphism of the algebra ${\cal F}$, for arbitrary
times $s$ and $t$, with 
\begin{equation*}
\alpha_{s,t}=\alpha_{s,t'}\circ\alpha_{t',t}.
\end{equation*}
\indent
Next, we describe two key hypotheses enabling us to study thermodynamic
processes, such as processes (I) and (II) described at the end of Section 1,
in the thermodynamic limit.\\

{\bf Hypothesis (A)}\ There is a class $\cal W$ of (time-dependent)
interactions s.t. the $*$automorphisms $\alpha_t^0$ and
$\alpha_{s,t}$, defined in \fer{51} and \fer{56}, with $W_t\in{\cal W}$,
satisfy the following property: for arbitrary $A\in{\cal F}$, 
\begin{equation}
n-\lim_{s\rightarrow\mp\infty}\alpha_s^0(\alpha_{s,0}(A)) =:\sigma_{\pm}(A)
\label{57}
\end{equation}
exists and defines a $*$endomorphism of ${\cal F}$.\\

{\bf Hypothesis (B)}\ For $H_{t_0}^\Lambda=\holambda_0$, the thermodynamic
limit of the equilibrium states $\varrho_{\beta,\bmu}$ defined in equation
\fer{15} exists on the $C^*$-algebra ${\cal F}$ and satisfies the KMS
condition (see e.g. \cite{2,5}), for arbitrary $\beta>0$, $\bmu\in\r^n$. \\

These two hypotheses can be verified in some simple, but physically relevant
examples; see Section 3 and \cite{5}. We now use them to discuss processes (I)
and (II). Let 
\begin{equation}
\omega^0=\omega_{\beta,\bmu}
\label{58}
\end{equation}
denote the state of ${\cal F}$ obtained as the thermodynamic limit of the
equilibrium states $\varrho_{\beta,\bmu}$ of equation \fer{15}. We are
interested in understanding the time dependence of the states 
\begin{equation}
\omega_t(A):=\omega^0(\alpha_{t_0,t}(A)), \ \ \ \ A\in {\cal F}.
\label{59}
\end{equation}
\indent
We first study this problem for {\bf process (I)}, with a limiting interaction 
$W_\infty\in{\cal W}$. Let $\alpha^\infty_{s,t}\equiv\alpha_{t-s}^\infty$
denote the $*$automorphism of ${\cal F}$ constructed in equation \fer{56} in
the example where $\holambda_t=\holambda_0+W_\infty$, for all $t$;
($\holambda_t$ is then time-independent, hence
$\alpha_{s,t}^\infty\equiv\alpha_{t-s}^\infty$ only depends on time
differences). We consider the operator
\begin{equation}
\e_A(t,s):=\alpha_{t_0,t}(\alpha_{s-t}^\infty(A)).
\label{60}
\end{equation}
The fundamental theorem of calculus yields
\begin{equation*}
\e_A(t,s)=\e_A(t_0,s) +\int_{t_0}^t{\e_A}'(u,s) du,
\end{equation*}
with 
\begin{equation}
{\e_A}'(u,s):=\frac{\partial}{\partial u}\e_A(u,s)=i\alpha_{t_0,u}\left(
[W_u-W_\infty, \alpha_{s-u}^\infty(A)]\right).
\label{62}
\end{equation}
Note that 
\begin{equation}
\|{\e_A}'(u,s)\|\leq 2\|A\|\, \|W_u-W_\infty\|,
\label{63}
\end{equation}
and the r.h.s. in \fer{63} tends to $0$, as $u\rightarrow\infty$, at an
integrable rate, see \fer{47}. Hence
\begin{equation}
\e_A(t,s)=\alpha_{s-t_0}^\infty(A)+i\int_{t_0}^t du\ \alpha_{t_0,u}\left(
  [W_u-W_\infty,\alpha_{s-u}^\infty(A)]\right).
\label{64}
\end{equation}
Since $\omega^0$ is invariant under the unperturbed time evolution
$\alpha_t^0$ (see equation \fer{51}, \fer{58}), it follows that
\begin{equation}
\omega^0(\e_A(t,s))=\omega^0(\alpha_{t_0-s}^0(\alpha_{s-t_0}^\infty(A))
+i\int_{t_0}^t du\ \omega_u\left([W_u-W_\infty,\alpha_{s-u}^\infty(A)]\right),
\label{65}
\end{equation}
see \fer{62}, \fer{59}, and the integral on the r.h.s. of \fer{65} converges
uniformly in $t$.\\
\indent
By Hypothesis (A), 
\begin{equation}
\lim_{s\rightarrow\infty}
\omega^0(\alpha_{t_0-s}^0(\alpha_{s-t_0}^\infty(A))=\omega^0(\sigma_+(A))
\label{66}
\end{equation}
exists, for arbitrary $A\in{\cal F}$. Furthermore, for arbitrary $u<\infty$, 
\begin{equation*}
\lim_{s\rightarrow\infty}
\omega_u\left([W_u-W_\infty,\alpha_{s-u}^\infty(A)]\right)=\lim_{s\rightarrow\infty}\omega_u\left([W_u-W_\infty,\alpha_{s-u}^0(\sigma_+(A))]\right),
\end{equation*}
again by Hypothesis (A), and it follows from the property of return to
equilibrium for the unperturbed time evolution, $\alpha^0_s$, that
\begin{equation}
\omega_u\left(
  [W_u-W_\infty,\alpha_{s-u}^0(\sigma_+(A))]\right)\rightarrow 0,
\label{67}
\end{equation}
as $s\rightarrow\infty$; see Section 3 for an example where return to
equilibrium holds for $\alpha_s^0$, and \cite{jp,5}. \\
\indent
In conclusion, the limit
\begin{equation}
\lim_{t\rightarrow\infty}
\omega^0(\alpha_{t_0,t}(A))=\lim_{t\rightarrow\infty}\omega^0 (\e_A(t,t))=\omega^0(\sigma_+(A)),\ \ A\in{\cal F},
\label{68}
\end{equation}
exists, by equations \fer{64}, \fer{66} and \fer{67}. It is known
from \cite{2} that the state $\omega^0(\sigma_+(\cdot))$ is an equilibrium
  (i.e. KMS) state for the asymptotic dynamics $\alpha_t^\infty$. \\
\indent
We note that equation \fer{68} is valid for any initial state $\omega^0$
which is invariant under $\alpha^0_t$ and has the property of return to
equilibrium, see also Section 3.
This completes our discussion of the thermodynamic process (I).\\ 
\indent
Next, we examine {\bf process (II)}. Since we
are interested in times $t\geq t_0$ we may consider the interaction $W_t$ to
be periodic with period $T$ for {\it all} times, $W_{t+T}=W_t$, for
$t\in\r$. It follows that
\begin{equation}
\alpha_{s,t}(A)=\alpha_{s+nT,t+nT}(A),
\label{69}
\end{equation}
for arbitrary times $s,t$, $n\in{\mathbb Z}$ and $A\in{\cal F}$. Decomposing
the time variable $t\in\r$ uniquely as $t=n(t)T+\tau(t)$, with $n(t)\in\mathbb
Z$ and $\tau(t)\in[0,T)$ we obtain from \fer{69} the equation
\begin{equation*}
\alpha_{t_0,t}(A)=\alpha_{t_0-n(t)T,0}\left(\alpha_{0,\tau(t)}(A)\right).
\end{equation*}
The invariance of $\omega^0$ under $\alpha_t^0$ and Hypothesis (A) then imply
that  
\begin{equation}
\lim_{t\rightarrow\infty}\left| \omega^0(\alpha_{t_0,t}(A))-\omega^0\left(\sigma_+\left(\alpha_{0,\tau(t)}(A)\right)\right)\right|=0,
\label{70}
\end{equation}
for all $A\in{\cal F}$. \\
\indent
The state given by
\begin{equation*}
\omega^P_t(A):= \omega^0\big(\sigma_+(\alpha_{0,\tau(t)}(A))\big),\ \
A\in{\cal F},
\end{equation*}
is periodic in $t$ with period $T$ (because $t\mapsto\tau(t)$ is). This shows
that $\omega_t$ approaches a {\it time-periodic} state as
$t\rightarrow\infty$. Notice that \fer{70} holds for an arbitrary
$\alpha_t^0$-invariant initial state $\omega^0$.\\

{\it Remark.\ } The approach to the asymptotic state (which is stationary for
Process (I) and time-periodic for Process (II)), for large times, holds for
arbitrary initial states which are normal w.r.t. the state 
$\omega^0$ given in \fer{58}. In other words, relations \fer{68} and \fer{70} hold if we replace
$\omega^0(\alpha_{t_0,t}(A))$ by $\omega(\alpha_{t_0,t}(A))$, for any state
$\omega$ on ${\cal F}$ which is normal w.r.t. $\omega^0$. The proof can be
found in \cite{5}.

\section{Thermodynamic processes for a reservoir of non-relativistic
  non-interacting fermions}

We consider an ideal quantum
gas of fermionic particles, e.g. modelling non-interacting, non-relativistic
electrons in a metal or a semi-conductor, subject to a time-dependent
perturbation. For the purpose of exposition, we concentrate here on spinless
fermions; a more general treatment can be found in \cite{4,5}. \\

The Hilbert space of pure states of the system confined to a bounded region
$\Lambda\subset\r^3$ is given by the fermionic Fock space over
$L^2(\Lambda,d^3x)$, 
\begin{equation}
\hslambda:= F_-(L^2(\Lambda,d^3x))=\bigoplus_{n\geq 0} P_-\left(
  L^2(\Lambda, d^3x)\right)^{\otimes n},
\label{71}
\end{equation}
where $P_-$ denotes the projection operator onto the subspace of antisymmetric 
functions, and where the subspace for $n=0$ is ${\mathbb C}$.\\ 
\indent
The non-interacting Hamiltonian is given by 
\begin{equation}
H_0^\Lambda:=  \bigoplus_{n\geq 0} h^\Lambda_n,
\label{72'}
\end{equation}
where $h^\Lambda_n$ acts on $P_-(L^2(\Lambda,d^3x))^{\otimes n}$ as
\begin{equation*}
h^\Lambda_n=\sum_{k=1}^n \bbbone\otimes\cdots\otimes\bbbone\otimes
(-\Delta)\otimes\bbbone\cdots\otimes\bbbone,
\end{equation*}
and $-\Delta$ is the Laplacian on $L^2(\Lambda,d^3x)$ with selfadjoint (e.g.
 Dirichlet-, Neumann-, or periodic) boundary conditions, acting on the $k$-th
 factor. We set $h^\Lambda_0=0$.\\ 
\indent
We define the field algebra as the CAR algebra (CAR for ``canonical
 anti-commutation relations'') 
\begin{equation*}
\Flambda :=\car\left(L^2(\Lambda,d^3x)\right),
\end{equation*}
which is the $C^*$-algebra generated by creation- and
annihilation operators,
\begin{equation*}
\{ a^\#(f)\ |\ f\in L^2(\Lambda,d^3x)\}.
\end{equation*}
The symbol $a^\#$ denotes either $a$ or $a^*$; recall that the
annihilation operator $a(f)$ acts on a wave-function $\psi\in
F_-(L^2(\Lambda,d^3x))$ as 
\begin{equation*}
(a(f)\psi)_n(x_1,\ldots,x_n)=\sqrt{n+1} \int \overline{f(x_{n+1})}
\psi_{n+1}(x_1,\ldots,x_n,x_{n+1}) d^3x_{n+1},
\end{equation*}
where $\psi_n$ is the projection of $\psi$ onto the $n$-particle subspace of
Fock space. The creation operators $a^*(f)$ (adjoint of $a(f)$) and
annihilation operators satisfy the canonical anti-commutation relations
\begin{equation*}
\begin{array}{c}
\{a(f),a^*(g)\}:=a(f)a^*(g)+a^*(g)a(f)=\int \overline{f(x)}g(x) d^3x\\
\{a^*(f),a^*(g)\}=\{a(f),a(g)\}=0,
\end{array}
\end{equation*}
for any $f,g\in L^2(\Lambda,d^3x)$. Notice that the $C^*$-algebra $\Flambda$
is weakly dense in ${\cal B}(\hslambda)$, 
\begin{equation*}
\left(\Flambda\right)''={\cal B}(\hslambda)
\end{equation*}
(double commutant). The non-interacting Hamiltonian \fer{72'} generates a $*$automorphism group $\alpha_t^{0,\Lambda}$ of $\Flambda$, according to
formula \fer{m2}, given by 
\begin{equation*}
\alpha_t^{0,\Lambda}\left(a^\#(f)\right)=a^\#(e^{-it\Delta}f),\ \ \ f\in
L^2(\Lambda, d^3x).
\end{equation*}
\indent
In this paper, we limit our discussion to only one
dynamical symmetry, namely the one corresponding to the
charge operator
\begin{equation}
N^\Lambda:=\bigoplus_{n\geq 0} n\bbbone\big|_{P_-(L^2(\Lambda,d^3x))^{\otimes
    n}}, 
\label{73}
\end{equation}
i.e., the particle number operator. 
We refer the reader to \cite{4,5} for a discussion involving more general
charges. It is obvious that the commutation relation \fer{11} is satisfied for
$\bQ^\Lambda=N^\Lambda$ and $\holambda_0$ given by \fer{73} and \fer{72'}, and
that the 
charge \fer{73} generates a $*$automorphism group $\phi_\tau^\Lambda$ on
$\Flambda$, 
according to \fer{12}. The observable algebra ${\cal A}^{\Lambda}$ defined in
\fer{22} corresponds to the $C^*$-algebra generated by monomials in creation-
and annihilation operators (smeared out with functions
in $L^2(\Lambda,d^3x)$) in which the number of
creation operators equals the number of annihilation operators. \\
\indent
We take the initial state of the system, at some fixed time $t_0$, to be given
by the density matrix
\begin{equation}
\varrho_{\beta,\mu}:=\left(\Xi_{\beta,\mu}^\Lambda\right)^{-1} \exp
-\beta\left[ H_0^\Lambda-\mu N^\Lambda\right].
\label{75}
\end{equation}
\indent
It is a standard result (see e.g. Section 5.2.4 of \cite{2}) that the dynamics
$\alpha_t^{0,\Lambda}$ has a thermodynamic limit, $\alpha_t^0$, in the sense
of equation 
\fer{51} and moreover, that the equilibrium state \fer{75} has a thermodynamic
limit in the sense that 
\begin{equation*}
\lim_{\Lambda\nearrow\r^3} \left(\Xi_{\beta,\mu}^\Lambda\right)^{-1}
\tr\left( \varrho_{\beta,\mu}A\right)=:\omega_{\beta,\mu}(A)
\end{equation*}
exists for all $A\in\car(L^2(\Lambda',d^3x))$ and any bounded
$\Lambda'\subset\r^3$ and defines an equilibrium (KMS) state at inverse
temperature 
$\beta$ and chemical potential $\mu$ on the $C^*$-algebra
\begin{equation}
{\cal F}=
\car(L^2(\r^3,d^3x))=\overline{\bigvee_{\Lambda\nearrow\r^3}
  \car(L^2(\Lambda,d^3x))}.
\label{77}
\end{equation}
This means that Hypothesis
(B) is verified.\\
\indent
It is known that the property of return to equilibrium holds for the KMS
state $\omega_{\beta,\mu}$ (relative to the dynamics
$\alpha_t^0$), see \cite{jp,1,5}; this means that 
\begin{equation*}
\lim_{t\rightarrow\pm\infty} \omega^0(B\alpha_t^0(A) C)=\omega^0(BC)\omega^0(A),
\end{equation*}
for all $A,B,C\in{\cal F}$. \\
\indent
So far, we have verified that our example is structurally compatible with the
theory outlined in Sections 1,2, and that the $\alpha_t^0$-invariant initial
state $\omega^0$ satisfies the property of return to equilibrium. We
are left with the specification of a class $\cal W$ of interactions satisfying
Hypothesis (A). \\
\indent
The description of such a class has been
given in \cite{5} for time-independent interactions, and we indicate here an extension to  time-dependent
ones. Set $x^{(N)}:=(x_1,\ldots,x_N)$, $x_j\in\r^3$ and
similarly for $y^{(N)}$, and let  
\begin{equation*}
w^N(t,x^{(N)}, y^{(N)})
\end{equation*}
be a function which is
bounded and continuously differentiable in $t\in\r$ and smooth and with
support in a compact region $\Lambda_0$ in each variable $x_j, y_j\in\r^3$. We denote by
\begin{equation*}
{\boldsymbol a}^*(x^{(N)}):= a^*(x_1)\cdots a^*(x_N)
\end{equation*} 
the product of creation operators $a^*(x_j)$ at positions $x_j \in\r^3$;
${\boldsymbol a}(y^{(N)})$ is defined 
similarly. The operator
\begin{equation}
W_t^N:=\int {\boldsymbol a}^*(x^{(N)})
w^N(t,x^{(N)},y^{(N)}) {\boldsymbol a}(y^{(N)})dx^{(N)} dy^{(N)},
\label{78}
\end{equation}
where we integrate over all spatial variables $x_j$ and $y_j$ in $\Lambda_0$,
defines 
an element of the $C^*$-algebra ${\cal F}$ introduced in \fer{77}. A norm-summable sequence of operators
$\{W_t^N\}_{N\geq 1}$ of the form \fer{78} determines an operator
\begin{equation}
W_t:=\sum_{N\geq 1}W_t^N\in{\cal F}.
\label{79}
\end{equation}
The class $\cal W$ consists of interactions of the form \fer{79} which satisfy
the smallness condition
\begin{equation*}
\|W_t\|_\infty' < \frac{1}{24\pi},
\end{equation*}
where we have introduced the norm
\begin{equation*}
\|W_t\|_\infty':=\sum_{N\geq 1}2^{5N} N \
\sup_{t\in\r}\|w^N(t,\cdot,\cdot)\|'_{6N}, 
\end{equation*}
with
\begin{equation*}
\|f\|'_M:=\frac{1}{2^{3M/2}}\left\langle f,\prod_{k=1}^M
  \left(-\frac{d^2}{dx_k^2}+x_k^2 +1\right)^3
  f\right\rangle^{1/2}_{L^2(\r^M)},
\end{equation*}
for a function $f\in L^2(\r^M)$ and where
$\langle\cdot,\cdot\rangle_{L^2(\r^M)}$ is the inner product in
$L^2(\r^M)$. \\
\indent
It is shown in \cite{5} that the limit \fer{57} exists, for time-independent
interactions in $\cal W$; the proof of convergence given there generalizes
readily to the time-dependent case, hence Hypothesis (A) holds. \\

{\it Remark.\ } One can explicitly calculate relevant physical quantities,
such as rates of change in time of internal energy, $\dot U(t)$, charge,
$\dot q(t)$, or 
entropy, $\dot S(t)$ (see \fer{40}, \fer{41}, \fer{44}) in the thermodynamic
limit in a {\it perturbative way} by using a Dyson series expansion
(which is norm-convergent uniformly in time); see equations
\fer{60}-\fer{68}. Furthermore, expectations in the reference state
$\varrho_t$ have a convergent perturbation expansion, as well. Thus, our
methods are quantitative.\\

{\bf Acknowledgements.\ } We thank Walter Kohn for attracting our interest to
problems of statistical mechanics involving time-periodic Hamiltonians and for
drawing our attention to reference \cite{A}.

\end{document}